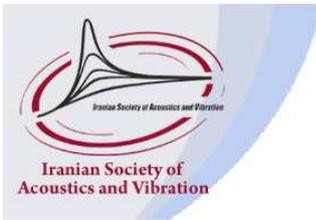 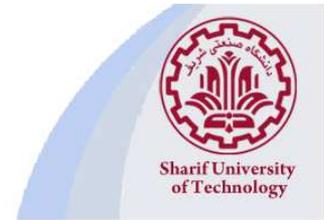

# Vibration transfer path analysis and path ranking for NVH optimization of a vehicle interior

B.Sakhaei[a], M.Durali[b], N.Hashemi[c]

[a] Deprtment of mechanical engineering, Sharif university of technology and NVH department of Iran Khodro Powertrain Company (IPCO), Theran, Iran

[b] Department of mechanical engineering, Sharif university of technology, Tehran, Iran

[c] Department of mechanical engineering, Amirkabir university of technology, Tehran, Iran

* Corresponding author e-mail: b_sakhaei@ip-co.com

## Abstract

By new advancements in vehicle manufacturing; evaluation of vehicle quality assurance has got a more critical issue. Today noise and vibration generated inside and outside the vehicles are more important factors for customers than previous. So far several researchers have focused on interior noise transfer path analysis and the results have been published in related papers but each method has its own limitations. In present work, the vibration transfer path analysis and vibration path ranking of a car interior has been performed. As interior vibration is a source of structural borne noise problem, thus the results of this research can be used to present the structural borne noise state in a vehicle. The method proposed in this paper, in opposite of the earlier methods, do not need to disassemble the power train from the chassis. The procedure shows a good ability of vibration path ranking in a vehicle and is an effective tool to diagnose the vibration problem inside the vehicle. The simulated vibration spectrums in different speeds of the engine have a good compliance with the tested results however some incompatibilities exist and have been discussed in details. The simulated results show the strength of the method in engine mount optimization.

## 1. Introduction

Noise and vibration which is perceived by passengers in a vehicle is important in the pleasantness of customers. Transfer path analysis of noise in a vehicle is a subject that many researchers from 90's up to now have worked on it to find the root cause of a noise problem in a vehicle. By taking help of these methods, the paths of noise, which usually starts from engine mounts through body and ends to passenger compartment, are investigated. Transfer path analysis can find the weak points of every path of the vehicle then the paths of high noise are identified and ranked. An NVH engineer then is able to find the problem and find a design solution for making the transmission of noise better.

The earliest work in this subject refers to Bendat [1] in 1980 which using coherence analysis of the noise paths to find various contributions. In late 80's an alternate method was proposed which consider the system as a source-transfer function-receiver and assumes that the noise in a vehicle compartment is a linear summation of different paths. In this method the response in target point in vehicle compartment is determined by multiplication of interface loading and transfer function from engine mount to that target point. Then noise contributions are summed to get the overall noise in the vehicle [2,3]. As it is clear the main challenge in this method is measuring the interface loads.[3]

Because of complexity in measuring the excitation forces on interfaces, the indirect procedures were developed. In these methods there is no need to direct measurement of interfacing forces. Instead FRFs between all points of source and the FRFs between source and receiver (targets) are measured. Then by inversing the FRF matrices and multiply it with the operational accelerations at the input side, the shares of each path from overall noise are calculated [4]. Although much advancement has promoted this method, and the accuracy of the results has been improved but indirect measurements have the limitations of cost and time of tests. In 2008 a new method based on operational modal analysis was proposed. This method named operational path analysis (OPA) which shortened the test time but it had accuracy problem because excitation in one direction often has side effect responses in other directions and putting this method in to effect needs high experience [5].

Multi level TPA has the strength of indirect measurement but can be done in a shorter time relative to indirect method. It was first introduced in 2002 by Eisele [6] et al which analyzed the interior structural noise in a vehicle. However the method is effective in prediction of the critical paths but less attention has been paid on it.

Although almost all publications in interior noise transfer path analysis has focused on interior noise simulation, there is no work on the interior vibration simulation. In this paper an interior vibration simulation of the vehicle for the first time was done. The method was based on multilevel TPA and the results show that this method has the ability of TPA analysis effectively. It is a fast method which rarely has the problem of measurement noise interfering. The results show that vehicle interior vibration simulation has good potential of engine mount optimization behaviour. By this method, also vibration fault diagnosis is more effective than conventional noise path ranking methods.

## 2. General formulation

Vibrations in a vehicle mostly transfer from engine mount locations through vibration transfer paths in to the body of the car and finally receive to the target locations in passenger compartment. Elements in vibration generation and transfer in to the vehicle are divided in to two major parts: active and passive elements. IC engine is an active vibration source and the engine mounts and body transfer functions from mount locations to target points in the vehicle compartment are passive elements.

In passive part of vibration transfer, each engine mount in each vehicle principal coordinate, comprise one path. Thus for a vehicle with 3 engine mounts, there are 9 transfer paths. These engine mount paths beside relative body and chassis transfer functions sends vibration energy to the passenger compartment.

Basic equation in transfer path analysis assumes that the total noise and vibration that feels at passenger position; is superposing the contribution of each path. Eq. (1) implies the relation: [4]

$$y(f) = \sum_{1}^{n} y_i(f) \tag{1}$$

Where

$y(f)$ = vibration at passenger location

$y_i(f)$ = contribution of vibration from each path

The system approach to the transfer path analysis explains that the partial contribution $y_i(f)$ of the vibration in the target point is a product of input force of the active part and the transfer function between the interface and the receiver like Eq. (2).

$$y_i(f) = FRF_{ik}(f) * F_i(f) \tag{2}$$

By combining the equations of (1) and (2), general basic equation of transfer path analysis is given by Eq. (3):

$$y(f) = \sum_{i=1}^{n} FRF_{ik}(f) * F_i(f) \quad (3)$$

As it is clear from the equation (3), it is assumed that vibration transfer paths have linear behaviour. Also it is obvious that transfer path analysis is performed in frequency region. In this method, as soon as any problem arises in over all amplitude of vibration, the different path contribution will be investigated and the responsible path for that problem will be identified. As each contribution is equal to the product of the input force and a transfer function then it is easier to locate the exact location of problem.

Multi level TPA is classified in fast TPA groups of methods in which the contribution to a target response is a chain of linked subsystems. In this method, few FRF measurements are being performed and then by multiplying the input signal to this chain, the output will be the vibration share of each path at target location. (Eq.(4))

$$\{y_i(f)\} = [H_1][H_2][H_3]\ldots\{F_i(f)\} \quad (4)$$

$F_i(f)$ = Input forces at engine mount interface

By noting the Eq.4, the basic equation of multilevel TPA can be written below

$$\{a_{interior}\} = \{F_i(f)\} \times \frac{a_{interior}}{a_{body}(f)} \rightarrow \frac{a_{interior}}{F_{body}(f)} = a_{engine}(f) \times \frac{a_{body}(f)}{a_{engine}(f)} \times \frac{F_{body}(f)}{a_{body}(f)} \times \frac{a_{interior}(f)}{F_{body}(f)} \quad (5)$$

According to Eq. (5), for evaluation of $a_{interior}$ from each path, it is needed to measure three transfer functions of mount transmissibility, apparent mass and chassis transfer function respectively.

## 3. Interior vibration simulation of a sedan car

For evaluation of interior vibration in a sedan car compartment a procedure of interior vibration simulation was applied based on multilevel TPA. The vehicle was equipped with a four cylinder engine of 1.7L. The engine and gearbox was installed on the chassis with three mounts, two rubber mounts and one hydraulic mount. The mounts are named as RH mount, LH mount and Rear mount. The RH mount was a hydraulic mount and LH and Rear mounts were rubber mounts. Figure (1) shows the transverse engine mounting system.[7]

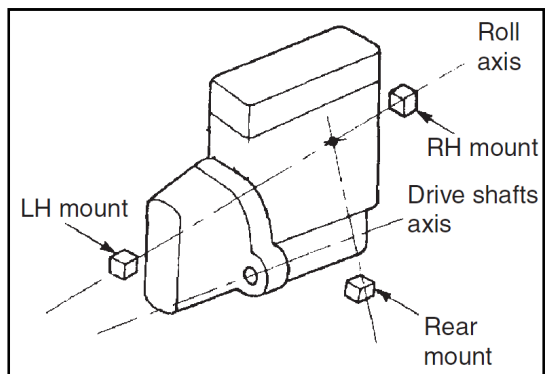

**Figure 1-** Transverse engine mounting system [7]

According to equation (5), two transfer functions of apparent mass and body were measured by presence of engine on the vehicle. In some publication the transfer function was measured without engine. As it was tested, (the results has not been reported) transfer functions without engine caused significant error in calculating the contributions.

All of the measurements were performed with a B&K 3570 data acquisition with 25Khz range FFT analyzer. There are two 4524 B&K triaxial accelerometers and a piezoelectric impact hammer of maximum 5KN force range. The signals were recorded with a 7 Hz high pass filter to prevent double hit error of impact hammer.

First the body transfer function ($a_{interior}(f)/F_{body}(f)$) was measured by exciting the engine mount location with an impact hammer. The impact was applied on the body side of engine mount and the force was measured with piezoelectric element of hammer. Simultaneously a triaxial accelerometer was installed on the vehicle compartment at passenger foot bottom on the floor. Figure (2)[7].As it was mentioned the excitation was applied at presence of engine and gearbox at original location. The frequency span of FFT the body transfer function was taken up to 800 Hz as only the vibrations of the vehicle interior were important. The frequency resolution of FFT analyzer was 0.25 Hz.

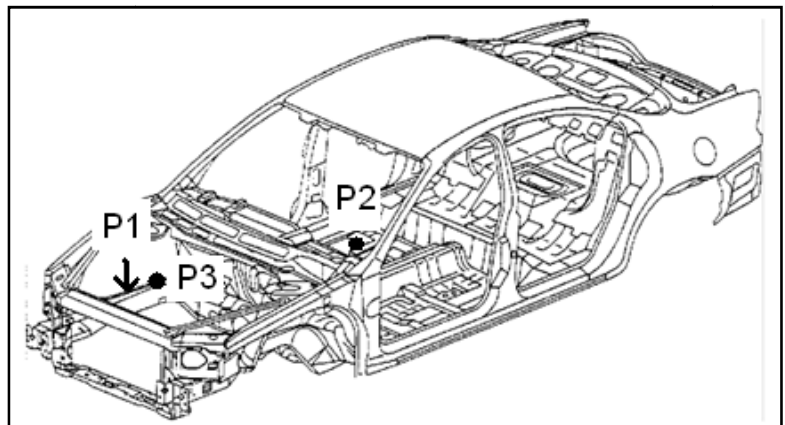

**Figure2-** P1:Hammer excitation, P2: Triax. acc. for body transfer func., P3: Triax. acc. for chassis apparent mass [7]

Figure (3) shows the body transfer function between RH mount location and vehicle compartment on passenger foot bottom at different principal directions.

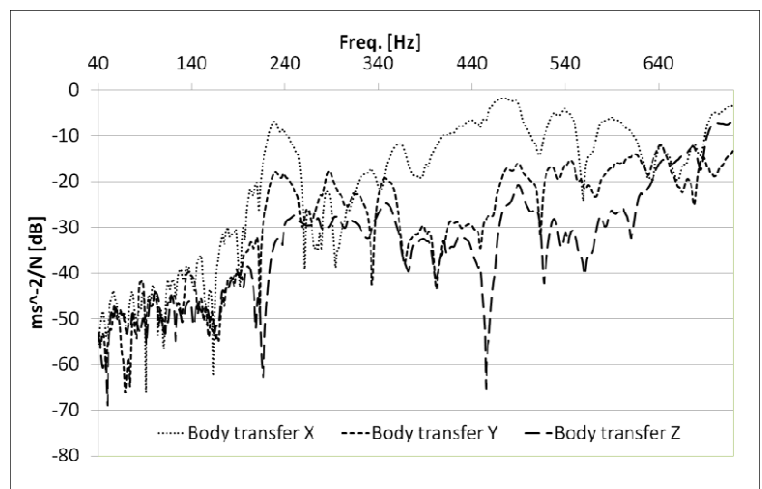

**Figure 3-** Body transfer function of RH mount at different directions

Apparent mass also was measured at engine mount locations. A trixial accelerometer was installed on body near the engine mount location of the vehicle and the impact hammer applied the force beside the accelerometer position.(Figure (2)). By calculating the ratio of $F_{body}(f)/a_{body}(f)$ an approximation of apparent mass could be got. (Figure (4))

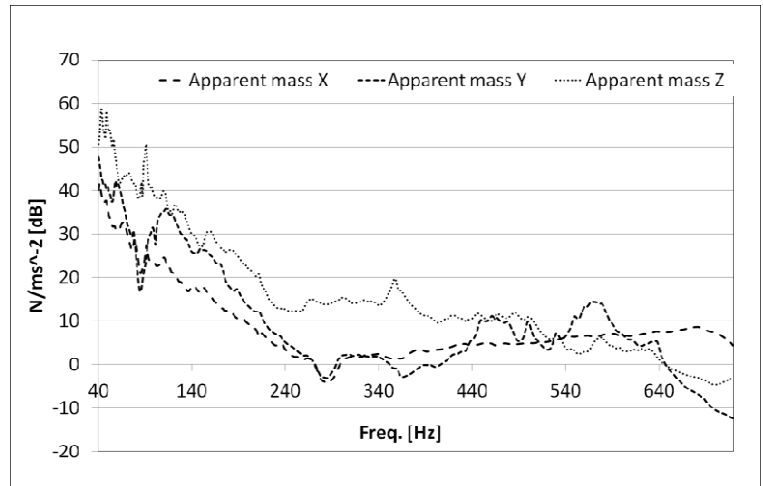

**Figure 4-** Apparent mass of RH mount region

Mount transmissibility is the ratio of acceleration on both sides of the mount on engine and body sides. The mount transmissibility was calculated by installing one accelerometer on engine side of each mount and another accelerometer on chassis side of mounts. Input and output acceleration on each engine mount were measured during a run up of engine speed on the chassis dynamometer. The 3$^{rd}$ gear was engaged during the test and engine was under full load condition (fully open throttle). Figure (5) shows the test set up to measure the mount transmissibility on chassis dyno.

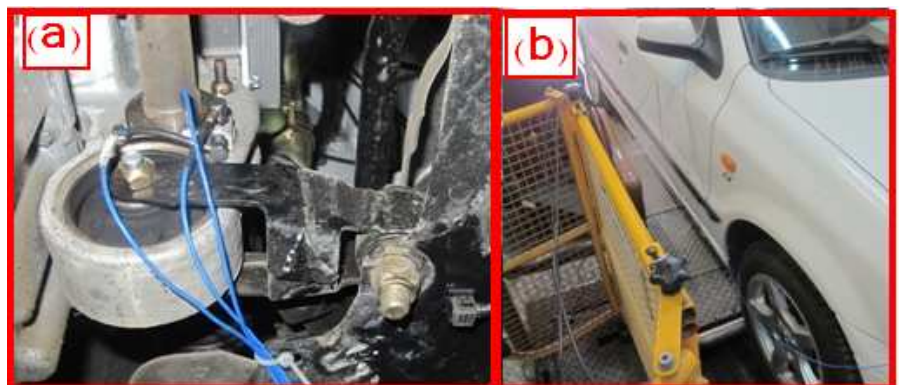

**Figure 5-** Test set up – a: Accelerometer installation on both sides of engine mount, b: Vehicle on chassis dyno.

The coherency spectrum of the mount transmissibility signal showed that only the main orders of the engine were coherent. Then only the transmissibility's of these main orders were taken in to account.

With a Matlab code the transmissibility of each mount at each direction was calculated at different engine speeds.

By multiplying the derived transfer functions with the input acceleration on engine side and summing the vibrations form different paths, the interior acceleration at passenger foot bottom could be simulated.

For comparison between simulated and real vibration at target point, a triaxial accelerometer also mounted at the target point (figure (2):P3) and simultaneously during the run up test, the vibration was measured. The results of the measurements are expressed in the next section.

## 4. Results

Calculation of mount dynamic stiffness in real conditions of mounts preload, temperature and under engine operation is one valuable benefit of multi level TPA. While dynamic mount stiffness measurement in test lab by a power shaker usually has large error.

Mount dynamic stiffness will be in hand by multiplying the mount transmissibility and apparent mass at different frequencies. Eq. (5).

Figure (6) shows the mount dynamic stiffness of RH mount. The mount dynamic stiffness decreases with frequency which complies with conventional engine mounts.[6]

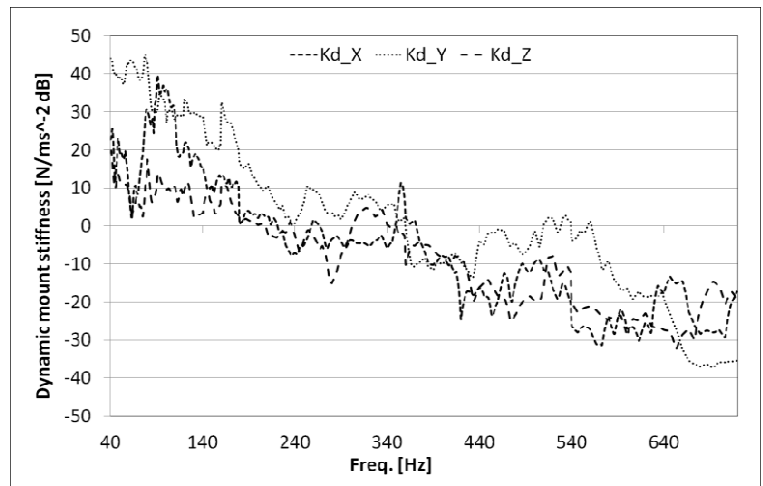

**Figure 6-** Dynamic mount stiffness derived by simulation

Figure (7) shows the comparison between simulated and measured overall accelerations at target point. There is a complete accordance between the trends of simulated and real signal but also still some differences exist.

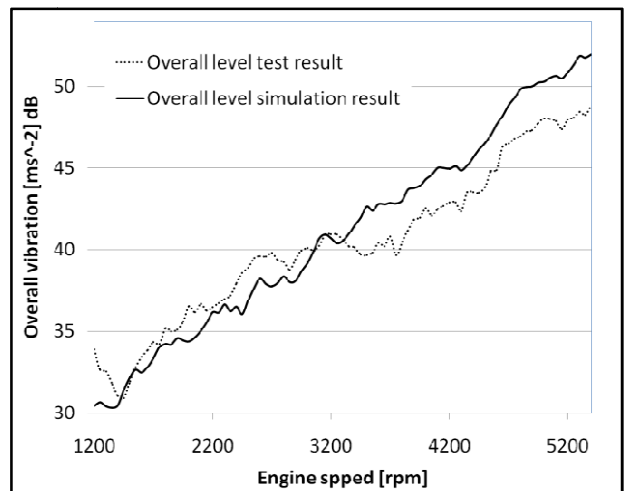

**Figure 7-** Dynamic mount stiffness derived by simulation

The differences between measured and simulated vibration signal in the vehicle comes from the damping of the materials which was covered the chassis of the vehicle on the foot bottom region. It was an asphalt layer. The property of damping of asphalt in this region is nonlinear which depends on temperature. As described earlier, the linear transfer function of body was taken in to account in the analysis then the damping nonlinear behaviour of this layer cannot be considered in the analysis.

The effect of damping also grows with the velocity of excitation. Therefore, at higher speeds the measured vibration curve in figure (5) was below the simulated.

These evidences shows that the damping of body transfer paths has nonlinear frequency dependent behaviour and this property should be studied with more care.

Figure (8) shows the contribution of each mount in overall vibration level. The LH mount has higher share of vibration in the compartment during engine run up especially from 1000 to 3500 rpm. Although RH mount behaviour in vibration transmission is desirable.

This shows that LH mount needs to be modified. Then softening of the LH mount can be a solution to this problem. Of course the side effect of mount softening on rigid body displacements of engine should be studied

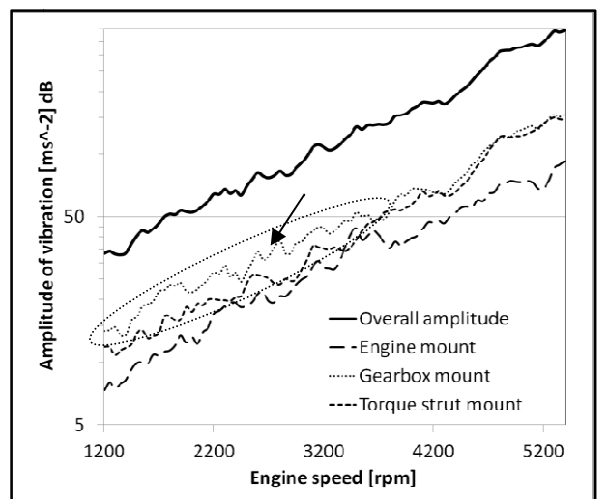

**Figure 8-** Dynamic mount stiffness obtained by simulation

## 5. Conclusion

Although there were many advancements in the transfer path analysis of noise in the vehicle, rare publications exists on vibration simulation of vehicle interior. Interior vibration of a vehicle can be taken as a representative of structural borne noise. Also vibration simulation of a vehicle interior is a powerful tool for engine mounts optimization. As it was proved, one can give applicable modifications on engine mount system to make the vibration behaviour better. Vibration TPA can realize the dynamic weak points of body chassis.

The results of vehicle interior vibration simulation and real measurements of this quantity showed a good compliance. The existed differences come from the nonlinear damping effect. Because of the linear assumption of body transfer functions in this method, these differences are inevitable.

A complete study on damping ratio effect on the simulated signal and also detailed mount optimization by this method will be performed in the next work. Also it is needed to further investigations be made on the effect different gears on the amount of vibration at target point.

## Acknowledgment

The authors are grateful to the IPCO (IRANKHODRO Power Train Co.) for supporting this research.

# REFERENCES

1. Bendat, J. S and Piersol, A.G., "Engineering Applications of Correlation and Spectral Analysis", *Wiley, New York*, 1980
2. Verheij, J., "Multipath Sound Transfer from Resiliently Mounted Shipboard Machinery", *PhD Dissertation*, Technische Physische Dienst TNO-TH, Delft, 1986
3. De Vis, D., Hendricx, W and Van Der Linden, P. "Development and Integration of an Advanced Unified Approach to Structure Borne Noise Analysis", *2nd Int. Conference on Vehicle Comfort*, ATA, 1992
4. Karl Janssens, Peter Mas, Ludo Gielen, Peter Gajdatsy and Herman van der Auweraer "A Novel Transfer Path Analysis Method Delivering a Fast and Accurate Noise Contribution Assessment", *SAE symposium on ineternational automotive technology*, SAE paper No. 2009-26-047
5. P. Gajdatsy, K.Janssens, WimDesmet, H.VanderAuweraer, "Application of the transmissibility concept in transfer path analysis", *Journal of mechanical systems and signal processing*, 2010
6. Norbert W. Alt, Norbert Wiehagen and Michael W. Schlitzer, "Interior Noise Simulation for Improved Vehicle Sound" *SAE noise and vibration conference and exposition*", SAE paper no.2001-01-1539
7. Jason C. Brown, A. John Robertson, Stan T. Serpento, "Motor vehicle structures: Concepts and fundamentals", Butter worth Heinemann, Oxford, 2002